# On the logarithmic profile of temperature in the atmospheric convective boundary layers

Yu Cheng[1]*, Qi Li[2], and Pierre Gentine[1]

[1]Department of Earth and Environmental Engineering, Columbia University, New York, NY 10027, USA

[2]School of Civil and Environmental Engineering, Cornell University, Ithaca, NY 14853, USA

*Email address for correspondence: yc2965@columbia.edu

**Abstract**

Wall-bounded turbulent flows are widely observed in natural and engineering systems, such as air flows near the Earth's surface, water flows in rivers, and flows around a car or a plane. The universal logarithmic velocity profile in wall-bounded turbulent flows proposed by von Kármán in 1930 is one of the few exact physical descriptions of turbulence. However, the mean velocity and temperature profiles cannot be adequately described by this universal log law when buoyancy effects are present. Monin-Obukhov similarity theory (MOST), proposed in 1954, has been the cornerstone theory to account for these buoyancy effects and to describe the atmospheric boundary layer. MOST has been used in almost all global weather, climate and hydrological models to describe the dependence of the mean velocity, temperature and scalar profiles on buoyancy. According to MOST, the logarithmic temperature profile breaks down as buoyancy effects become important. In contrast, here we show that this long-standing MOST theory does not apply for temperature. We propose a new theory for the logarithmic profile of near-wall temperature, which corrects MOST pitfalls and is supported by both high-resolution direct numerical simulations and field observations of the convective atmospheric boundary layer. Buoyancy effects do not modify the logarithmic nature but instead modulate the slope of the temperature profile compared to the universal von Kármán slope. The new formulation has widespread applications such as in climate models, where the proposed new temperature log law should lead to more realistic continental surface temperature, which are strongly impacted by buoyancy.

**Introduction**

The log law of velocity (1) in wall-bounded turbulent flows is one of the cornerstones (2, 3) of turbulence theory. Similarly, the log law for mean temperature is widely known (4, 5) to apply in wall-bounded turbulent flows, where buoyancy effects are absent and when temperature can be treated as a passive scalar. The near-wall temperature and velocity profiles and turbulent fluxes in the atmospheric surface layer (around the lowest 10% of the ABL (6)) are the key boundary conditions for numerical weather prediction (7), global climate models (8-10) and hydrological models (11). Monin-Obukhov similarity theory (MOST), developed in 1954, aims at correcting the log law in the presence of buoyancy effects (12) and has been the cornerstone of atmospheric boundary layer turbulence. MOST has since been used in most applications to define atmospheric boundary layer fluxes (7-10) as well as the temperature and velocity profiles near the surface and how they are modified with varying degrees of buoyancy. MOST (12) corrects the log profiles of wind and scalars using a dimensional analysis, based on stability and distance to the wall $z$, assuming that the logarithmic profile needs to be corrected as instability increases (details in equation [9]). Yet, it has been recently demonstrated that MOST (12) does not take into account the outer layer scaling, such as the depth of the boundary layer $z_i$, which however could be important for surface layer flows (13). We now have the capacity to accurately simulate and observe the atmospheric surface layer so that MOST can be reevaluated and systematically tested.

**Results**

To obtain the profiles of temperature and velocity in the near-wall region, direct numerical simulations (DNS) of convective boundary layers ranging from weakly unstable (convective) to highly unstable (13) and free convection (14) are conducted. The three simulations of convective boundary layer flow (13) named Sh20, Sh5 and Sh2 are forced with varying mean geostrophic wind. The stability parameter $z_i/L$ varies between $-7.1$, $-105.1$ and $-678.2$ (from weakly to highly convective) in those simulations, respectively, where $z_i$ is the convective boundary layer height and $L$ is the Obukhov length (15), with a constant and uniform flux boundary condition at the surface. Another simulation of free convection named Microhh ReL (similar to van Heerwaarden and Mellado (14)), uses a constant temperature boundary condition at $\frac{z_i}{L} = -100\,288.4$. Details of the four DNS experiments can be found in the Materials and Methods section and are summarized in Supplementary Information Table S1.

Similarly to Kader and Yaglom (4), the difference between the mean ($x$-$y$ plane horizontal average) potential temperature $\Theta$ at each height $z$ and the mean potential temperature $\Theta_h$ at the lowest DNS grid is normalized by a scaling temperature $\theta_*$ (Materials and Methods). The instantaneous dimensionless temperature $\frac{\Theta-\Theta_h}{\theta_*}$ fits a log law in $z^+$ across all DNS datasets (Fig. 1), where $z^+ = \frac{z u_\tau}{\nu}$ is the normalized height $z$ in inner units, $u_\tau$ is the friction velocity and $\nu$ the kinematic viscosity. The coefficient of determination $R^2$ for $\frac{\Theta-\Theta_h}{\theta_*}$ and $\log(z^+)$ is 1.00 for the selected vertical zone near the wall across the DNS datasets, emphasizing that the temperature follows a log law across conditions. However, the normalized velocity $\frac{U}{u_\tau}$, with $U$ the mean streamwise velocity, does not follow such a log law (details in Supplementary Information Fig. S4) in more convective conditions ($R^2 \leq 0.22$ at $\frac{z_i}{L} \leq -105.1$). The deviation from velocity log law is due to buoyancy effects (16-19), which is also suggested by MOST. Such a distinct behavior of temperature and velocity has also been recently reported in turbulent natural convection (20).

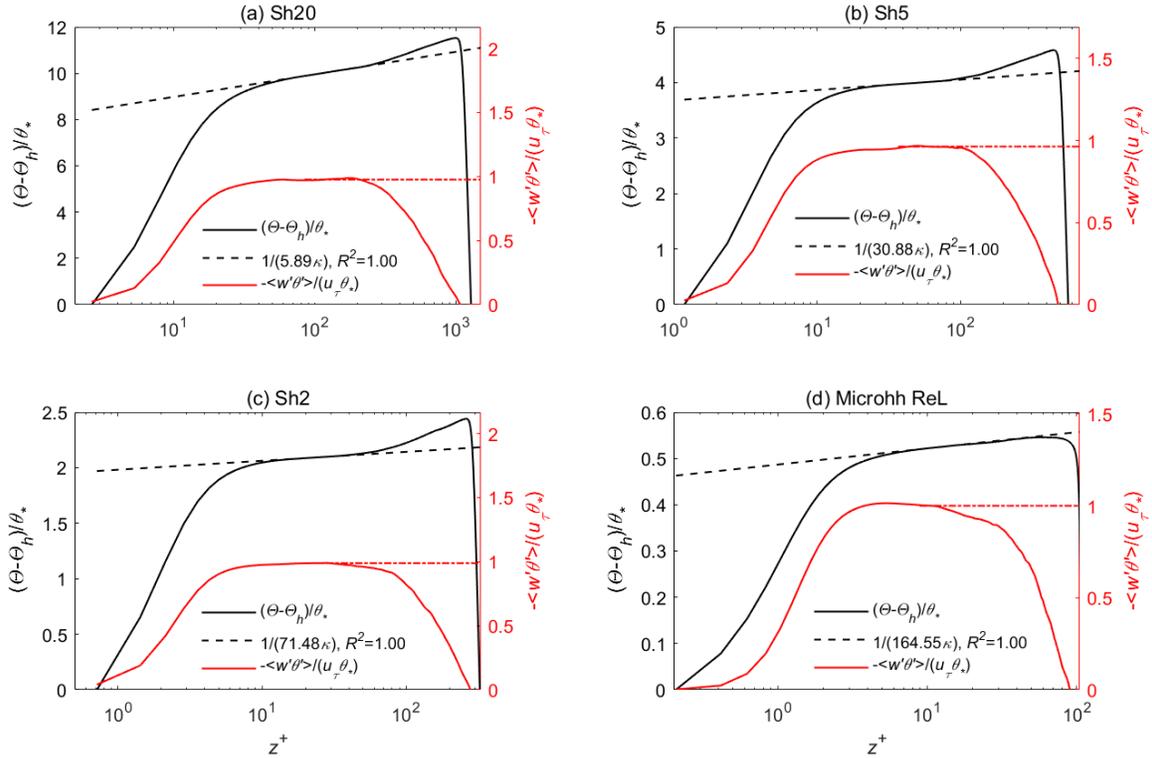

**Figure 1.** Vertical profiles of the normalized potential temperature and heat flux averaged in the $x$-$y$ plane in different convective DNS datasets. Details of the DNS datasets (**a**) Sh20, (**b**) Sh5, (**c**) Sh2 and (**d**) Microhh ReL are introduced in Supplementary Information Table S1. $\Theta$ is mean potential temperature in the $x$-$y$ plane at each height, $\Theta_h$ is mean potential temperature in the $x$-$y$ plane at the lowest DNS grid, $\theta_*$ is a scaling temperature, $z^+ = \frac{z u_\tau}{\nu}$ is normalized vertical coordinate, $u_\tau$ is the friction velocity, $\nu$ is the kinematic viscosity, $\kappa$ is the von Kármán constant, $R^2$ is the coefficient of determination, $w'$ is the fluctuation of vertical velocity, $\theta'$ is the fluctuation of potential temperature and $<>$ denotes averaging in the $x$-$y$ plane. The black dashed line denotes the fitted log profile and the slope is shown. The red dashed line denotes the mean heat flux in the constant heat flux zone.

The coexistence of a temperature log law and constant heat flux observed in the DNS datasets resembles the coexistence of the velocity log law and constant momentum flux in turbulent shear flows (21). This constant heat flux zone is similar to the atmospheric surface layer, which by definition has nearly constant fluxes (6). The black dashed line in Fig. 4 is used to denote more precisely the vertical zone where the temperature log law and constant heat flux coexist. The slope of the DNS temperature log law is not constant but instead decreases from $\frac{1}{5.89\kappa}$ to $\frac{1}{164.55\kappa}$ when $z_i/L$ decreases from $-7.1$ to $-100\,288.4$, where $\kappa \approx 0.40$ is the von Kármán constant. This is in contrast with the universal log law for mean velocity in turbulent shear flows (22) which has a constant slope of $\frac{1}{\kappa}$. Such a variation of the temperature slope was also observed in recent Rayleigh-Bénard convection (23). In the vertical region where the temperature log law exists, the turbulent heat flux $\overline{w'\theta'}$ is almost constant across the DNS datasets (Fig. 1), whether a constant heat flux boundary condition is applied like in simulations Sh20, Sh5 and Sh2 or whether a constant surface temperature is prescribed like in Microhh ReL.

To provide theoretical foundation for our log law observation for potential temperature in the presence of buoyancy, we turn to the transport equation for the heat flux $\overline{w'\theta'}$ in the horizontally homogeneous atmospheric boundary layer (6)

$$\frac{\partial \overline{w'\theta'}}{\partial t} = \frac{g}{\overline{\theta}}\overline{\theta'\theta'} - \overline{w'w'}\frac{\partial \overline{\theta}}{\partial z} - \frac{\partial \overline{\theta'w'w'}}{\partial z} - \frac{1}{\overline{\rho}}\overline{\theta'\frac{\partial p'}{\partial z}}, \qquad [1]$$

where $t$ is time, $g$ the gravitational acceleration, $\overline{\theta}$ the mean potential temperature (also denoted as $\Theta$), $\overline{\rho}$ the mean density, $p'$ the pressure fluctuation and $^-$ (or $<>$) the horizontal averaging in the $x$-$y$ plane. With some assumptions, shown in the Materials and Methods section, the $\overline{w'\theta'}$ transport equation can be simplified to

$$-\frac{\overline{w'\theta'}}{\frac{\partial \overline{\theta}}{\partial z}} = \frac{u_\tau \theta_*}{\frac{\partial \overline{\theta}}{\partial z}} = 3.6\alpha_3\alpha_4 \left(\frac{z_i}{L}\right)^{\frac{2}{3}} \frac{u_\tau^2}{w_*} z, \qquad [2]$$

where $\alpha_3$ is independent of $z$ and is a constant in the stability range $-678.2 \leq \frac{z_i}{L} \leq -7.1$ based on the DNS datasets, $\alpha_4$ is a parameter that may vary with $z$ and $\frac{z_i}{L}$, and $w_*$ is a convective velocity scale (24). If $\alpha_4$ does not vary with $z$ in the constant heat flux zone, the above equation suggests that $-\frac{\overline{w'\theta'}}{\frac{\partial \overline{\theta}}{\partial z}}$ should be linearly related to $z$.

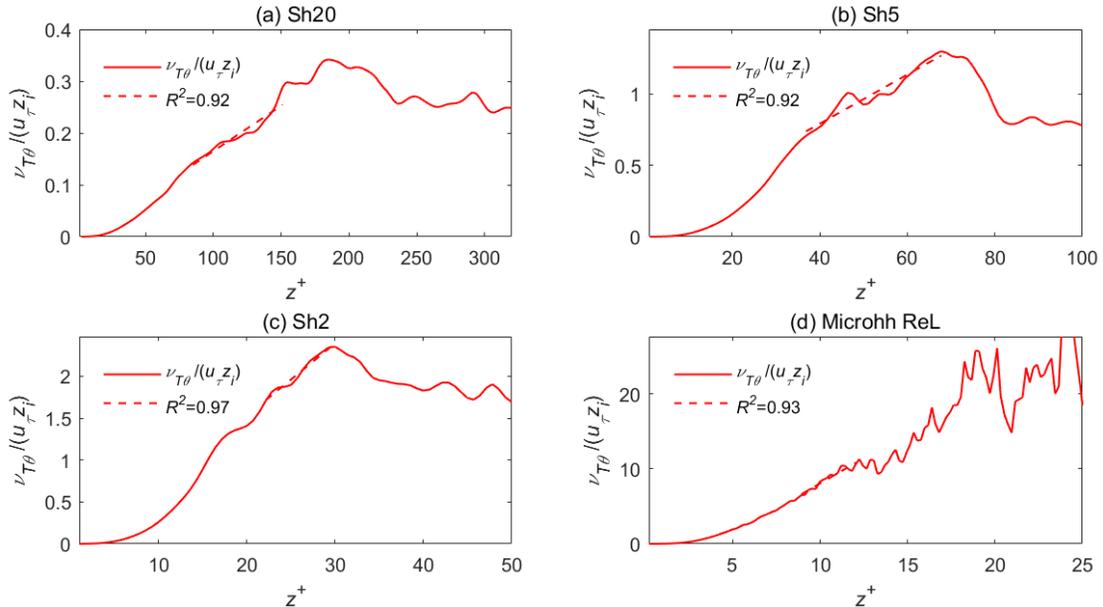

**Figure 2.** The dimensionless turbulent eddy diffusivity $\frac{\nu_{T\theta}}{u_\tau z_i}$ for potential temperature in the $z$ direction in different convective DNS data. $\nu_{T\theta}$ is the turbulent diffusivity for potential temperature and other variables have the same definition as those in Fig. 1. The red dashed line indicates linear regression in the constant heat flux zone.

We further define a turbulent eddy viscosity coefficient for potential temperature as $\nu_{T\theta} \equiv -\frac{\overline{w'\theta'}}{\frac{\partial \overline{\theta}}{\partial z}}$. The DNS datasets support our theoretical derivation and the fact that $\nu_{T\theta} \propto z$ (Fig. 2), thus $\alpha_4$ can be regarded independent of $z$ in the constant heat flux layer across convective conditions in the DNS experiments. We further define a variable $\kappa_\theta$ as below

$$\kappa_\theta \equiv \frac{\nu_{T\theta}}{u_\tau z} = 3.6 \alpha_3 \alpha_4 \frac{u_\tau}{w_*} \left(\frac{z_i}{L}\right)^{2/3}. \tag{3}$$

Based on our analysis in Materials and Methods, $\kappa_\theta$ does not depend on $z$ since both $\alpha_3$ and $\alpha_4$ do not vary with $z$ in the constant heat flux region. The $\overline{w'\theta'}$ transport equation can then be written in a form similar to the neutral velocity log law $\frac{u_\tau \theta_*}{\frac{\partial \overline{\theta}}{\partial z}} = \kappa_\theta u_\tau z$, so that the potential temperature profile with buoyancy is rewritten as:

$$\frac{\overline{\theta} - \overline{\theta}_0}{\theta_*} = \frac{1}{\kappa_\theta} \log\left(\frac{z}{\delta}\right) + B, \tag{4}$$

where $\overline{\theta}_0$ is mean potential temperature at the wall, $\delta = \nu/u_\tau$ is the viscous length scale, $B$ is a parameter independent of $z$. This constitutes our new theory for the profile of potential temperature in the convective boundary layers ranging from the weakly unstable condition to free convection. We note that the conditions $\nu_{T\theta} \propto z$ and $\kappa_\theta$ independent of $z$ are required for the existence of the temperature log law. The slope of the temperature log law is $\frac{1}{\kappa_\theta}$, which is a modulation of the von Kármán constant used in the absence of buoyancy effects. The ratio of $\kappa_\theta$ to the von Kármán constant $\kappa$ is $\frac{\kappa_\theta}{\kappa} = \frac{3.6}{\kappa} \alpha_3 \alpha_4 \frac{u_\tau}{w_*} \left(\frac{z_i}{L}\right)^{2/3}$. The DNS datasets suggest that $\alpha_4$ is not dependent on $z$, but one may wonder how the increased convection influences the potential temperature profile. The convective boundary layer height can be written as $z_i = \frac{w_*^3}{\frac{g}{\overline{\theta}} \overline{w'\theta'}}$ according to the definition of $w_*$ (24). Taking the ratio of $z_i$ and the Obukhov length $L$, we obtain

$$\frac{w_*}{u_\tau} = \left(-\frac{z_i}{L}\frac{1}{\kappa}\right)^{1/3}. \tag{5}$$

Then the ratio of $\kappa_\theta$ and the von Kármán constant $\kappa$ can be rewritten as a function of the boundary layer height $z_i$

$$\frac{\kappa_\theta}{\kappa} = \frac{3.6}{\kappa^{1/3}} \alpha_3 \alpha_4 \frac{w_*}{u_\tau} = \frac{3.6}{\kappa^{2/3}} \alpha_3 \alpha_4 \left(\frac{z_i}{-L}\right)^{1/3}. \tag{6}$$

The DNS datasets show that $\frac{\kappa_\theta}{\kappa} \propto \left(\frac{w_*}{u_\tau}\right)^2 = \left(\frac{z_i}{-L}\frac{1}{\kappa}\right)^{\frac{1}{3}*2}$ (Fig. 3a) and thus $\alpha_4 \propto \left(\frac{z_i}{-L}\right)^{1/3}$ in our stability conditions $\frac{z_i}{L}$ ranging from $-678.2$ to $-7.1$ since $\alpha_3$ is a constant. In the neutral limit though, $\frac{z_i}{-L} = 1$, the temperature log law is expected to have the same slope as the velocity log law in neutral turbulent shear flows, i.e., $\frac{\kappa_\theta}{\kappa} = 1$. Thus, we obtain

$$\frac{\kappa_\theta}{\kappa} = \left(\frac{z_i}{L}\right)^{2/3} = \kappa^{2/3} \left(\frac{w_*}{u_\tau}\right)^2, \tag{7}$$

for a larger range of stabilities than our DNS, ranging from $-678.2$ to $-1$. Besides, the DNS datasets suggest that another coefficient $B$ in the temperature log law equation satisfies the relation $B \propto \left(\frac{w_*}{u_\tau}\right)^{-1.03} = \left(\frac{z_i}{-L}\frac{1}{\kappa}\right)^{\frac{-1.03}{3}}$ in the stability range $-678.2 \leq \frac{z_i}{L} \leq -7.1$ (Fig. 3b). Further analyses from the DNS datasets suggest that $B \approx 20 \left(\frac{w_*}{u_\tau}\right)^{-1} = 20 \left(\frac{z_i}{-L}\frac{1}{\kappa}\right)^{-\frac{1}{3}}$ in the stability range $-678.2 \leq \frac{z_i}{L} \leq -7.1$ but it is not as precisely determined as the slope $\frac{1}{\kappa_\theta}$ in log laws (4).

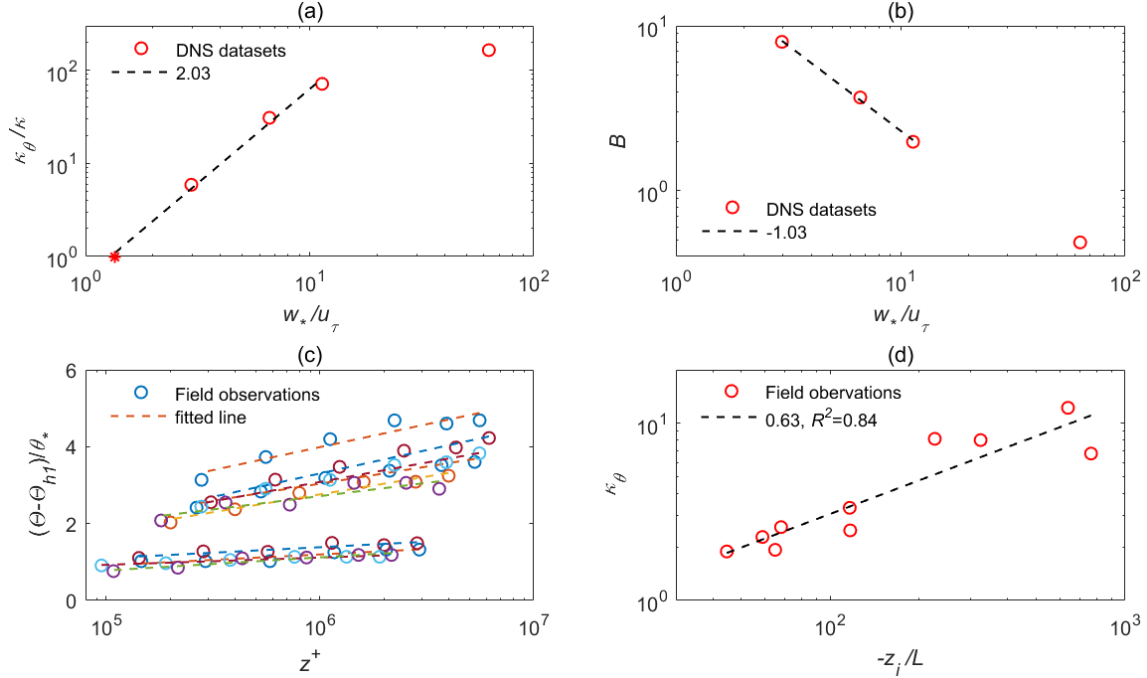

**Figure 3.** Parameters of the temperature log law profile determined from the DNS datasets and field observations, and the normalized potential temperature in field observations of the convective atmospheric boundary layer. (**a**) The ratio $\frac{\kappa_\theta}{\kappa}$ and (**b**) the temperature log law coefficient $B$ plotted against $\frac{w_*}{u_\tau}$ in various convective DNS data. The red asterisk in (a) denotes the case when $\frac{\kappa_\theta}{\kappa}=1$ in the neutral limit $\frac{z_i}{-L}=1$. (**c**) Vertical profiles of normalized potential temperature in convective conditions in the Cabauw Experimental Site for Atmospheric Research in the Netherlands. The vertical potential temperature profiles in 10 different 30-minute convective periods in July 2019 are shown. $\theta$ is the averaged potential temperature at heights of 10 m, 20 m, 40 m, 80 m, 140 m and 200 m above the land surface on a tower in a 30-minute period, and $\theta_{h1}$ is the averaged potential temperature at the height 2 m in a 30-minute period. (**d**) The inverse of the temperature log law slope denoted by $\kappa_\theta$ is plotted against $\frac{z_i}{-L}$ in various convective conditions shown in (c). The fitted slopes are shown in (a), (b) and (d) and the coefficient of determination $R^2$ is shown in (d). The coefficient of determination $R^2$ for all 10 periods in (c) are above 0.80.

In addition to the DNS experiments, we analyze field observations (details in Materials and Methods) of the convective atmospheric boundary layer in the Cabauw Experimental Site for Atmospheric Research (4.926° E, 51.97° N) in the Netherlands, which are at higher Reynold number than the DNS experiments. A linear relation is fitted between the dimensionless temperatures $\frac{\theta-\theta_{h1}}{\theta_*}$ and $\log(z^+)$ in the selected 10 periods (Fig. 3c), where $\theta$ is the averaged potential temperature at heights of 10 m, 20 m, 40 m, 80 m, 140 m and 200 m above the land surface on a tower in a 30-minute period and $\theta_{h1}$ is the averaged potential temperature at the height 2 m in a 30-minute period. Moreover, a linear relation can also be fitted between the inverse of the slope of the temperature log law (denoted by $\kappa_\theta$) and $z^+$ in log-log plot, with an $R^2=0.84$ (Fig. 3d). The field observations confirms our DNS findings and the variations of the logarithmic slope with stability, $\kappa_\theta \propto \left(\frac{-z_i}{L}\right)^{0.63}$, which closely matches the relation $\kappa_\theta \propto \left(\frac{-z_i}{L}\right)^{2/3}$ from the DNS datasets. Beside the caveat that the observation of the boundary layer height $z_i$ is limited by resolution (25, 26), this observational match gives us confidence that our DNS results are universal, and not impacted by the (higher) Reynolds number observable in the atmospheric boundary layer.

**Discussion**

The proposed temperature log law in the convective boundary layers can be written as $\frac{\kappa_\theta z}{\theta_*}\frac{\partial \theta}{\partial z}=1$, or equivalently

$$\frac{\kappa z}{\theta_*}\frac{\partial \theta}{\partial z} = f\left(\frac{z_i}{L}\right), \qquad [8]$$

where $f\left(\frac{z_i}{L}\right) \equiv \frac{\kappa}{\kappa_\theta}$ is a function that can be approximated as $f\left(\frac{z_i}{L}\right) = \left(\frac{z_i}{L}\right)^{-2/3}$ in the stability range $-678.2 \leq \frac{z_i}{L} \leq -1$. According to MOST, the mean potential temperature was instead assumed to depend on $z/L$ (12)

$$\frac{\kappa z}{\theta_*}\frac{\partial \theta}{\partial z} = \phi_h\left(\frac{z}{L}\right), \qquad [9]$$

where $\phi_h$ is a stability correction function dependent on the distance to the wall – leading to a non-logarithmic profile. Both $f\left(\frac{z_i}{L}\right)$ and $\phi_h\left(\frac{z}{L}\right)$ are corrections of the neutral temperature equation $\frac{\kappa z}{\theta_*}\frac{\partial \theta}{\partial z} = 1$. In our new derivation and in our observations, the function $f\left(\frac{z_i}{L}\right)$ does not depend on $z$ thus leading to a log law. Instead, such a dependence on $z$ of $\phi_h\left(\frac{z}{L}\right)$, was assumed in MOST based on dimensional analysis with $z$, which is now shown to be incorrect. In our various convective DNS datasets, $\frac{\kappa z}{\theta_*}\frac{\partial \theta}{\partial z}$ approaches a constant that is equal to $\frac{\kappa}{\kappa_\theta}$ in the constant heat flux zone (Fig. 4), thus supporting a log law for temperature rather than MOST. In our new theory, the proposed log layer depends on an outer layer scaling, the boundary layer height $z_i$. This outer layer correction, $f\left(\frac{z_i}{L}\right)$, is consistent with recent studies emphasizing the importance of outer layer scaling $z_i$ compared to the distance to the wall $z$ in convective conditions (27-30).

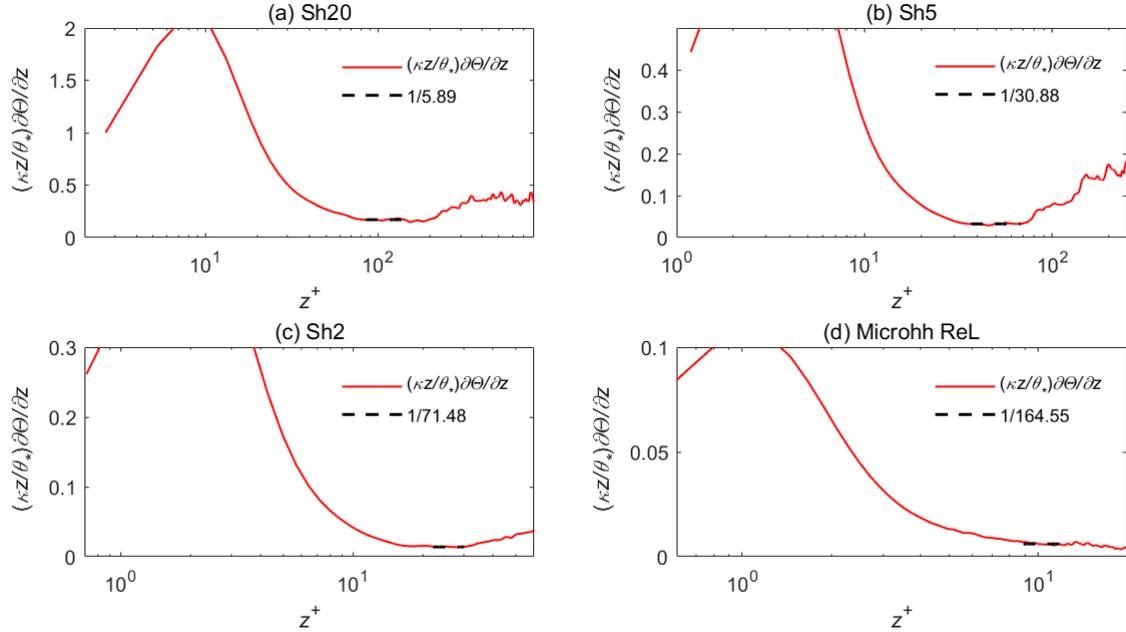

**Figure 4.** The dimensionless temperature gradient $\frac{\kappa z}{\theta_*}\frac{\partial \theta}{\partial z}$ in the $z$ direction in different convective DNS data. All variables have the same definition as those in Fig. 1. The black dashed line denotes the average $\frac{\kappa z}{\theta_*}\frac{\partial \theta}{\partial z}$ in the constant heat flux zone.

We report a new theory for the potential temperature profile in the near-wall region affected by buoyancy effects, through DNS and field observations of the convective boundary layers ranging from the weakly convective condition to free convection. The new temperature log law can be described by $\frac{\kappa_\theta z}{\theta_*}\frac{\partial \theta}{\partial z} = 1$, where $\kappa_\theta = \kappa \left(\frac{z_i}{L}\right)^{2/3}$ is valid in the stability range $-678.2 \leq \frac{z_i}{L} \leq -1$. We suggest applying the proposed temperature log profile to global climate models and wall models for large eddy simulations where MOST is generally applied.

**Materials and Methods**

**DNS of convective boundary layers**. The convective ABL has been studied extensively using large eddy simulations (LESs) (31-33). However, uncertainties exist with their subgrid-scale models near the wall (13, 34) and wall-modeled LESs for atmospheric studies are often based on Monin-Obukhov similarity theory (32, 35-37). Recently, DNS have been used to study the convective ABL (38-41), which can resolve the full range of turbulence scales, although the DNS Reynolds number is smaller than that in the atmosphere. In the 3 convective simulations named Sh2, Sh5 and Sh20, the incompressible Navier-Stokes equations with Boussinesq approximation are solved using the code described in a previous study (13). The boundary conditions for the temperature field are constant flux at the surface and zero flux at the top of the computational domain. Periodic boundary conditions are employed in the horizontal ($x$ and $y$) directions. The grid points for the dataset Sh2 are $nx \times ny \times nz = 1200 \times 800 \times 602$, while the grid points for both Sh5 and Sh20 are $1200 \times 800 \times 626$ in streamwise ($x$), spanwise ($y$) and vertical ($z$) directions, respectively. The Reynolds number is defined as $Re_\tau = \frac{u_\tau z_i}{\nu}$, where $u_\tau$ is the friction velocity and $\nu$ the kinematic viscosity. Details of the DNS setup can be found in previous studies (13, 42) and are summarized in Supplementary Information Table S1. The selected time step is when the horizontally averaged potential temperature profile is almost in steady state.

Details of another simulation of free convection named Microhh ReL (the simulation ReL in Heerwaarden and Mellado (14)) can be found in previous studies (14, 43) and are summarized in Supplementary Information Table S1. The boundary condition for the potential temperature field is constant temperature at the surface and constant temperature gradient at the top boundary. The grid points are $nx \times ny \times nz = 1536 \times 1536 \times 768$. Reynolds number similarity has been observed in the simulation thus the DNS results may be extrapolated to higher Reynolds numbers in the ABL (14). The selected time step is 404 s when the vertically integrated kinetic energy does not vary much with time (14).

**Field observations of the convective atmospheric boundary layer**. The Cabauw Experimental Site for Atmospheric Research (44) (http://www.cesar-observatory.nl/) has a tower of 213 m in height with observations at 2 m, 10 m, 20 m, 40 m, 80 m, 140 m and 200 m above a grass land in the Netherlands (4.926° E, 51.97° N), thus providing unique muiti-level temperauture observations in the atmospheric boundary layer. A number of 30-minute periods at 11:00-15:00 UTC in July 2019 are used as the raw data.

The boundary layer height is retrieved from Lufft CHM 15k ceilometer (45), which is used to detect the top of an elevated aerosol layer. The ceilometer backscatter profiles can be used to retrieve the ABL height for convective conditions when the ABL is well mixed and there are significant differences between the aerosol content of the ABL and the free troposphere (46). The average ABL height of each 30-minute period is used as the raw data.

The potential temperature is retrieved from validated temperature measurements at heights of 2 m, 10 m, 20 m, 40 m, 80 m, 140 m and 200 m above the land surface on the 213-meter-high tower. The validated temperature data is named cesar_tower_meteo_lb1_t10_v1.2_201907.nc and can be downloaded from http://www.cesar-database.nl. The average potential temperature of each 30-minute period is used as the raw data.

The validated surface fluxes is named cesar_surface_flux_lb1_t10_v1.0_201907.nc and is downloaded from http://www.cesar-database.nl. The average validated surface fluxes of each 30-minute period is used as the raw data.

The dimensionless temperatures $\frac{\theta - \theta_{h1}}{\theta_*}$ and $\log(z^+)$ at heights of 10 m, 20 m, 40 m, 80 m, 140 m and 200 m are fitted by a linear relation. The raw data are filtered base on two criteria: the coefficient of determination $R^2 > 0.80$ (for $\frac{\theta - \theta_{h1}}{\theta_*}$ and $\log(z^+)$); and the boundary layer height is larger than 1100 m. In fact, $R^2 > 0.80$ is found in more than half of the temperature profiles of the 30-minute periods in the daytime. The boundary layer height restriction is based on the rough estimation that the atmospheric surface layer, i.e., the constant flux layer, is approximately the lowest 10% of the ABL (6). Besides, we would like to include the measurements at 200 m to ensure a wider zone of the log law and to keep more available ceilometer-observed ABL heights. Note that other restrictions on the boundary layer height can lead to different selected periods but the fitted slope between $\frac{\theta - \theta_{h1}}{\theta_*}$ and $\log(z^+)$ does not vary too much. These two criteria leave 10 different 30-minute periods in July 2019.

**Detailed derivation of the temperature log law.** The horizontally averaged potential temperature equation in the convective boundary layers can be written as

$$\frac{\partial \bar{\theta}}{\partial t} = \nu_\theta \frac{\partial^2 \bar{\theta}}{\partial z^2} - \frac{\partial \overline{w'\theta'}}{\partial z}, \qquad [10]$$

where $\theta'$ is the fluctuation from mean potential temperature $\bar{\theta}$ (also denoted as $\Theta$), $t$ the time coordinate, $\nu_\theta$ the thermal diffusivity, $z$ the vertical coordinate, $w'$ is the vertical velocity fluctuation and $^{-}$ the horizontal averaging in the $x$-$y$ plane. Assuming steady state, i.e., $\frac{\partial \bar{\theta}}{\partial t} = 0$, the above equation is reduced to

$$\nu_\theta \frac{\partial^2 \bar{\theta}}{\partial z^2} - \frac{\partial \overline{w'\theta'}}{\partial z} = 0. \qquad [11]$$

Integrating the above equation from 0 to some height $z$ near the wall, we have

$$\nu_\theta \frac{\partial \bar{\theta}}{\partial z}\bigg|_{z=0} = \nu_\theta \frac{\partial \bar{\theta}}{\partial z} - \overline{w'\theta'}. \qquad [12]$$

In the constant heat flux zone near the wall, neglecting the thermal diffusivity term we obtain (6)

$$u_\tau \theta_* \equiv \nu_\theta \frac{\partial \bar{\theta}}{\partial z}\bigg|_{z=0} = -\overline{w'\theta'}, \qquad [13]$$

where $u_\tau$ is the friction velocity and $\theta_*$ the temperature scaling.

The transport equation for the heat flux $\overline{w'\theta'}$ in the convective atmospheric boundary layer can be written in the form (6)

$$\frac{\partial \overline{w'\theta'}}{\partial t} = \frac{g}{\bar{\theta}}\overline{\theta'\theta'} - \overline{w'w'}\frac{\partial \bar{\theta}}{\partial z} - \frac{\partial \overline{\theta'w'w'}}{\partial z} - \frac{1}{\rho}\overline{\theta'\frac{\partial p'}{\partial z}}. \qquad [14]$$

On the right-hand side of the above equation are the buoyancy term, shear term, transfer term and pressure term from left to right, respectively. By definition, the Obukhov length (15) is $L = -\frac{u_\tau^3}{\frac{\kappa g}{\bar{\theta}}\overline{w'\theta'}}$. The ratio of the transfer term to the buoyancy term is

$$\gamma_1 \equiv \frac{-\frac{\partial \overline{\theta'w'w'}}{\partial z}}{\frac{g}{\bar{\theta}}\overline{\theta'\theta'}}. \qquad [15]$$

The DNS datasets suggest that $\gamma_1$ does not change sign with $z$ in the constant heat flux zone at convective conditions (Supplementary Information Fig. S1). The vertically averaged $\gamma_1$ in the constant heat flux zone varies from 0.18 to 0.24 in the stability range $-678.2 \leq \frac{z_i}{L} \leq -7.1$, and approaches $-0.06$ at $\frac{z_i}{L} = -100\,288.4$.

The pressure term in $\overline{w'\theta'}$ equation is approximated by the Rotta model (47) while keeping the slow component and the buoyancy term (48),

$$\frac{1}{\rho}\overline{\theta'\frac{\partial p'}{\partial z}} \approx \frac{\overline{w'\theta'}}{\tau_\theta} + \alpha_1 \frac{g}{\bar{\theta}}\overline{\theta'\theta'}, \qquad [16]$$

where $\tau_\theta$ is a return-to-isotropy time scale and $\alpha_1$ is a constant (48, 49). Assuming steady state and multiplying each term by $\frac{\kappa z}{u_\tau^2 \theta_*}$ as in Wyngaard, Coté and Izumi (50), the $\overline{w'\theta'}$ transport equation in the constant heat flux zone can be simplified as

$$\frac{\overline{w'w'}}{u_\tau^2}\frac{\kappa z}{\theta_*}\frac{\partial \bar{\theta}}{\partial z} - (1 - \alpha_1 + \gamma_1)\frac{z}{L}\frac{\overline{\theta'\theta'}}{\theta_*^2} - \frac{\kappa z}{u_\tau \tau_\theta} = 0. \qquad [17]$$

The ratio of the buoyancy term and the pressure term is

$$\alpha_2 \equiv \frac{\frac{g}{\bar{\theta}}\overline{\theta'\theta'}}{\frac{1}{\rho}\overline{\theta'\frac{\partial p'}{\partial z}}} \approx \frac{\frac{g}{\bar{\theta}}\overline{\theta'\theta'}}{\frac{\overline{w'\theta'}}{\tau_\theta} + \alpha_1 \frac{g}{\bar{\theta}}\overline{\theta'\theta'}}. \qquad [18]$$

The DNS datasets show that the vertically averaged $\alpha_2$ varies from 0.64 at weakly convective conditions ($\frac{z_i}{L} = -7.1$) to 2.66 at free convection ($\frac{z_i}{L} = -100288.4$). Besides, the variation of $\alpha_2$ with height $z$ is less than 11% in the constant heat flux zone within each DNS experiment. Therefore, $\alpha_2$ is mainly a function of $\frac{z_i}{L}$.

According to the definition of $\alpha_2$ and the Rotta model, we obtain

$$\frac{\frac{\overline{w'\theta'}}{\tau_\theta}}{\frac{g}{\bar{\theta}}\overline{\theta'\theta'}} = \frac{1}{\alpha_2} - \alpha_1. \qquad [19]$$

The $\overline{w'\theta'}$ transport equation can then be reduced to

$$\frac{u_\tau \theta_*}{\frac{\partial \bar{\theta}}{\partial z}} = \frac{\kappa z \frac{\overline{w'w'}}{u_\tau}}{\frac{1 - \alpha_2 - \alpha_2 \gamma_1}{1 - \alpha_1 \alpha_2}\frac{\kappa z}{u_\tau \tau_\theta}} = \frac{1 - \alpha_1 \alpha_2}{1 - \alpha_2 - \alpha_2 \gamma_1}\overline{w'w'}\tau_\theta. \qquad [20]$$

Similarly to the definition of the velocity relaxation time $\tau_u \equiv \frac{e}{\epsilon}$ (47, 51), we have

$$\tau_\theta \equiv \frac{\overline{\theta'\theta'}}{2\epsilon_\theta}, \qquad [21]$$

where $e$ is the turbulent kinetic energy (TKE), $\epsilon$ is the TKE dissipation rate and $\epsilon_\theta$ is the dissipation rate of potential temperature variance. The DNS datasets suggest that $\tau_\theta \propto z^{1/3}$ (Supplementary Information Fig. S2) can be a good approximation in the constant heat flux zone at various convective conditions.

Above the Obukhov length $-L$, heat flux and turbulent intensities are functions only of $\frac{z}{z_i}$ and $w_*$, which is defined as (24, 52)

$$w_* = \left(\frac{g}{\bar\theta} z_i \overline{w'\theta'}\right)^{1/3}. \qquad [22]$$

Moeng and Wyngaard (48) showed that $\tau_\theta$ is also related to boundary layer height $z_i$. From dimensional analysis and the DNS results, we obtain

$$\tau_\theta = \alpha_3 \frac{z_i^{\frac{2}{3}} z^{\frac{1}{3}}}{w_*}, \qquad [23]$$

where $\alpha_3 \approx 0.36$ in the stability range $-678.2 \leq \frac{z_i}{L} \leq -7.1$ and $\alpha_3$ decreases to 0.21 at $\frac{z_i}{L} = -100288.4$. The $\overline{w'\theta'}$ transport equation can be further reduced to

$$\frac{u_\tau \theta_*}{\frac{\partial \bar\theta}{\partial z}} = \frac{1-\alpha_1\alpha_2}{1-\alpha_2-\alpha_2\gamma_1} \alpha_3 \overline{w'w'} \frac{z_i^{\frac{2}{3}} z^{\frac{1}{3}}}{w_*}. \qquad [24]$$

The group of coefficients in the above equation can be written as

$$\alpha_4 \equiv \frac{1-\alpha_1\alpha_2}{1-\alpha_2-\alpha_2\gamma_1}, \qquad [25]$$

which may vary with $\frac{z_i}{L}$ in the stability range $-678.2 \leq \frac{z_i}{L} \leq -7.1$ mainly due to the variation of $\alpha_2$. And $\alpha_4$ might also vary height $z$ in the constant heat flux region due to the variations of $\gamma_1$ and $\alpha_2$. At highly convective conditions, Wyngaard, Coté and Izumi (50) obtained the following approximation

$$\overline{w'w'} = 3.6 u_\tau^2 \left(\frac{z}{L}\right)^{2/3}, \qquad [26]$$

which is also a good approximation for the constant heat flux zone in the DNS data (Supplementary Information Fig. S3). As it becomes more unstable, $\overline{w'w'}$ becomes closer to $(z^+)^{2/3}$ in the constant heat flux zone. The $\overline{w'\theta'}$ transport equation can thus be written as

$$-\frac{\overline{w'\theta'}}{\frac{\partial \bar\theta}{\partial z}} = \frac{u_\tau \theta_*}{\frac{\partial \bar\theta}{\partial z}} = 3.6 \alpha_3 \alpha_4 \left(\frac{z_i}{L}\right)^{\frac{2}{3}} \frac{u_\tau^2}{w_*} z. \qquad [27]$$

**Supplementary Information**

**Velocity profiles in the near-wall region**. A log law for normalized velocity $\frac{U}{u_\tau}$ and $z^+$ is fitted using the same slope as the temperature log profile in the constant heat flux zone denoted by the blue dashed line (Supplementary Information Fig. S4), where $U$ is the mean streamwise velocity. The coefficient of determination $R^2$ is 1.00 in the weakly unstable case Sh20 ($\frac{z_i}{L} = -7.1$) and the variation of momentum flux $\overline{w'u'}$ is only 6% in the constant heat flux zone. Therefore, a velocity log law and constant momentum flux can still be observed in the weakly unstable condition ($\frac{z_i}{L} = -7.1$) with a slope $\frac{1}{5.89\kappa}$ smaller than $\frac{1}{\kappa}$ of turbulent shear flows. However, the coefficient of determination $R^2$ decreases to 0.22 at more unstable conditions ($\frac{z_i}{L} = -105.1$ and $\frac{z_i}{L} = -678.2$) and even to $-0.03$ at free convection. Meanwhile, the variation of the momentum flux $\overline{w'u'}$ is 17% ($\frac{z_i}{L} = -105.1$), 11% ($\frac{z_i}{L} = -678.2$) and 76% ($\frac{z_i}{L} = -100288.4$) in the constant heat flux zone, respectively. Therefore, the velocity log law or constant momentum flux is not observed in more convective conditions ($\frac{z_i}{L} \leq -105.1$), which is still consistent with the deviation from the velocity log law due to buoyancy effects (16-19).

The horizontally averaged velocity equation in the convective boundary layers can be written as

$$\frac{\partial \bar u}{\partial t} = \nu \frac{\partial^2 \bar u}{\partial z^2} - \frac{\partial \overline{w'u'}}{\partial z}, \qquad [28]$$

where $u'$ is the fluctuation from mean streamwise velocity $\bar u$ (also denoted as $U$). Assuming steady state, i.e., $\frac{\partial \bar u}{\partial t} = 0$, the above equation is reduced to

$$\nu \frac{\partial^2 \bar{u}}{\partial z^2} - \frac{\partial \overline{w'u'}}{\partial z} = 0. \qquad [29]$$

Integrating the above equation from 0 to some height $z$ near the wall, we have

$$\nu \frac{\partial \bar{u}}{\partial z}\bigg|_{z=0} = \nu \frac{\partial \bar{u}}{\partial z} - \overline{w'u'}. \qquad [30]$$

In the constant momentum flux zone near the wall, neglecting the viscous term we obtain

$$u_\tau^2 \equiv \nu \frac{\partial \bar{u}}{\partial z}\bigg|_{z=0} = -\overline{w'u'}. \qquad [31]$$

The transport equation for the momentum flux $\overline{w'u'}$ in the convective atmospheric boundary layer can be written in the form (6)

$$\frac{\partial \overline{w'u'}}{\partial t} = \frac{g}{\bar{\theta}}\overline{u'\theta'} - \overline{w'w'}\frac{\partial \bar{u}}{\partial z} - \frac{\partial \overline{u'w'w'}}{\partial z} - \frac{1}{\bar{\rho}}\overline{u'\frac{\partial p'}{\partial z} + w'\frac{\partial p'}{\partial x}}. \qquad [32]$$

On the right-hand side of the above $\overline{w'u'}$ equation are the buoyancy term, shear term, transfer term and pressure term from left to right, respectively. The ratio of the transfer term to the pressure term in $\overline{w'u'}$ equation is

$$\gamma_{11} \equiv \frac{\frac{\partial \overline{u'w'w'}}{\partial z}}{\frac{1}{\bar{\rho}}\overline{u'\frac{\partial p'}{\partial z} + w'\frac{\partial p'}{\partial x}}}. \qquad [33]$$

The DNS datasets suggest that $\gamma_{11}$ changes sign with $z$ in the constant heat flux zone in the stability range $-100\,288.4 \leq \frac{z_i}{L} \leq -105.1$ (Supplementary Information Fig. S5). The mean of $\gamma_{11}$ in the constant heat flux region varies between $-0.24$ and $0.66$ in the stability range $-678.2 \leq \frac{z_i}{L} \leq -7.1$, and approaches $-23.26$ at $\frac{z_i}{L} = -100288.4$.

The pressure decorrelation term is approximated by the Rotta model (47) while keeping the slow component and the buoyancy term (48),

$$\frac{1}{\bar{\rho}}\overline{u'\frac{\partial p'}{\partial z} + w'\frac{\partial p'}{\partial x}} \approx \frac{\overline{w'u'}}{\tau_u} + \alpha_1 \frac{g}{\bar{\theta}}\overline{u'\theta'}, \qquad [34]$$

where $\tau_u$ is a return-to-isotropy time scale and $\alpha_1$ is a constant (48, 49). The ratio of the buoyancy term to the pressure term is

$$\alpha_{22} \equiv \frac{\frac{g}{\bar{\theta}}\overline{u'\theta'}}{\frac{1}{\bar{\rho}}\overline{u'\frac{\partial p'}{\partial z} + w'\frac{\partial p'}{\partial x}}} \approx \frac{\frac{g}{\bar{\theta}}\overline{u'\theta'}}{\frac{\overline{w'u'}}{\tau_u} + \alpha_1 \frac{g}{\bar{\theta}}\overline{u'\theta'}}. \qquad [35]$$

The DNS datasets show that the ratio of the buoyancy term to pressure term does not change sign with $z$ in the stability range $-678.2 \leq \frac{z_i}{L} \leq -7.1$ but changes sign with $z$ at $\frac{z_i}{L} = -100288.4$ in the constant heat flux layer. The mean of $\alpha_{22}$ in the constant heat flux layer decreases from $0.57$ to $0.33$ when $\frac{z_i}{L}$ decreases from $-7.1$ to $-678.2$. At $\frac{z_i}{L} = -100288.4$, the mean of $\alpha_{22}$ is $7.70$ in the constant heat flux zone.

The $\overline{w'u'}$ transport equation can then be reduced to

$$\alpha_1(\alpha_{22} - \gamma_{11} - 1)\frac{z}{L}\frac{\overline{u'\theta'}}{\overline{w'\theta'}} + \frac{\overline{w'w'}}{u_\tau^2}\frac{\kappa z}{u_\tau}\frac{\partial \bar{u}}{\partial z} + \frac{\kappa z}{u_\tau \tau_u}\frac{\overline{w'u'}}{u_\tau^2}(1 + \gamma_{11} - \alpha_{22}) = 0, \qquad [36]$$

Assuming constant momentum flux at some region near the wall, we have

$$-\alpha_1(1 + \gamma_{11} - \alpha_{22})\frac{z}{L}\frac{\overline{u'\theta'}}{\overline{w'\theta'}} + \frac{\overline{w'w'}}{u_\tau^2}\frac{\kappa z}{u_\tau}\frac{\partial \bar{u}}{\partial z} - \frac{\kappa z}{u_\tau \tau_u}(1 + \gamma_{11} - \alpha_{22}) = 0. \qquad [37]$$

After some algebra, we obtain

$$-\frac{\alpha_1\alpha_{22}(1+\gamma_{11}-\alpha_{22})}{1-\alpha_1\alpha_{22}}\frac{\kappa z}{u_\tau \tau_u} + \frac{\overline{w'w'}}{u_\tau^2}\frac{\kappa z}{u_\tau}\frac{\partial \bar{u}}{\partial z} - \frac{\kappa z}{u_\tau \tau_u}(1+\gamma_{11}-\alpha_{22}) = 0, \qquad [38]$$

which can be further reduced to

$$\frac{u_\tau^2}{\frac{\partial \bar{u}}{\partial z}} = \frac{1-\alpha_1\alpha_{22}}{1+\gamma_{11}-\alpha_{22}}\overline{w'w'}\tau_u. \qquad [39]$$

Similarly to $\tau_\theta$, the approximation $\tau_u \propto z^{1/3}$ can be applied in the DNS datasets (Supplementary Information Fig. S6), although the approximation is not as good as that for $\tau_\theta$. In analogy with $\tau_\theta$, from dimensional analysis and the DNS results, we obtain

$$\tau_u = \alpha_{33}\frac{z_i^{\frac{2}{3}}z^{\frac{1}{3}}}{w_*}, \qquad [40]$$

where $\alpha_{33} \approx 1.7$ in the stability range $-678.2 \leq \frac{z_i}{L} \leq -7.1$ and $\alpha_3$ increases to 3.5 at $\frac{z_i}{L} = -100288.4$. Again we apply the approximation $\overline{w'w'} = 3.6 u_\tau^2 \left(\frac{z}{L}\right)^{2/3}$ of Wyngaard, Coté and Izumi (50) as in the derivation of temperature log law. Thus we obtain

$$-\frac{\overline{w'u'}}{\frac{\partial \bar{u}}{\partial z}} = \frac{u_\tau^2}{\frac{\partial \bar{u}}{\partial z}} = 3.6\frac{1-\alpha_1\alpha_{22}}{1+\gamma_{11}-\alpha_{22}}\alpha_{33}\left(\frac{z_i}{L}\right)^{\frac{2}{3}}\frac{u_\tau^2}{w_*}z. \qquad [41]$$

The coefficients in the above equation can be written as

$$\alpha_{44} \equiv \frac{1-\alpha_1\alpha_{22}}{1+\gamma_{11}-\alpha_{22}}. \qquad [42]$$

$\alpha_{44}$ may vary with $z$ due to the variation of $\gamma_{11}$ with $z$ in the stability range $-678.2 \leq \frac{z_i}{L} \leq -105.1$. Invoking the eddy viscosity assumption, we have

$$\nu_T \equiv -\frac{\overline{w'u'}}{\frac{\partial \bar{u}}{\partial z}}, \qquad [43]$$

where $\nu_T$ is the turbulent viscosity. The DNS datasets support that $\nu_T \propto z$ (Supplementary Information Fig. S7) at the weakly unstable condition ($\frac{z_i}{L} = -7.1$). However, $\nu_T$ oscillates a lot with $z$ at and is not linearly related to $z$ at $\frac{z_i}{L} = -105.1$, thus supporting the variation of $\alpha_{44}$ with $z$. Moreover, there are negative values of $\nu_T$ in the near wall region at $\frac{z_i}{L} \leq -105.1$ in contrast to positive $\nu_{T\theta}$ (Fig. 2), which may further influence the possible log law. Therefore, there is not a general velocity log law in convective conditions, which is also consistent with previous descriptions of velocity profiles (16).

**Table S1.** Details of DNS set-up of convective boundary layers ranging from the weakly convective condition to free convection. $Re_\tau = \frac{u_\tau z_i}{\nu}$ is Reynolds number, $u_\tau$ is the friction velocity, $\nu$ is the kinematic viscosity, $z_i$ is the boundary layer height, $L$ is the Obukhov length, $L_x$, $L_y$ and $L_z$ are the domain sizes in the $x$, $y$ and $z$ directions, respectively. $\Delta_x^+ = \frac{\Delta_x u_\tau}{\nu}$, $\Delta_y^+$ and $\Delta_z^+$ are the spatial grid resolutions denoted by inner units in the $x$, $y$ and $z$ directions, respectively.

| DNS data | $Re_\tau$ | $\frac{z_i}{L}$ | $\frac{L_x}{L_y}$ | $\frac{L_x}{L_z}$ | $\Delta_x^+$ | $\Delta_y^+$ | $\Delta_z^+$ |
|---|---|---|---|---|---|---|---|
| Microhh ReL | 80 | -100288.4 | 1 | 1.87 | 0.19 | 0.19 | 0.21 |
| Sh2 | 309 | -678.2 | 1.5 | 6 | 2.87 | 2.87 | 0.71 |
| Sh5 | 554 | -105.1 | 1.5 | 6 | 4.95 | 4.95 | 1.19 |
| Sh20 | 1243 | -7.1 | 1.5 | 6 | 11.02 | 11.02 | 2.65 |

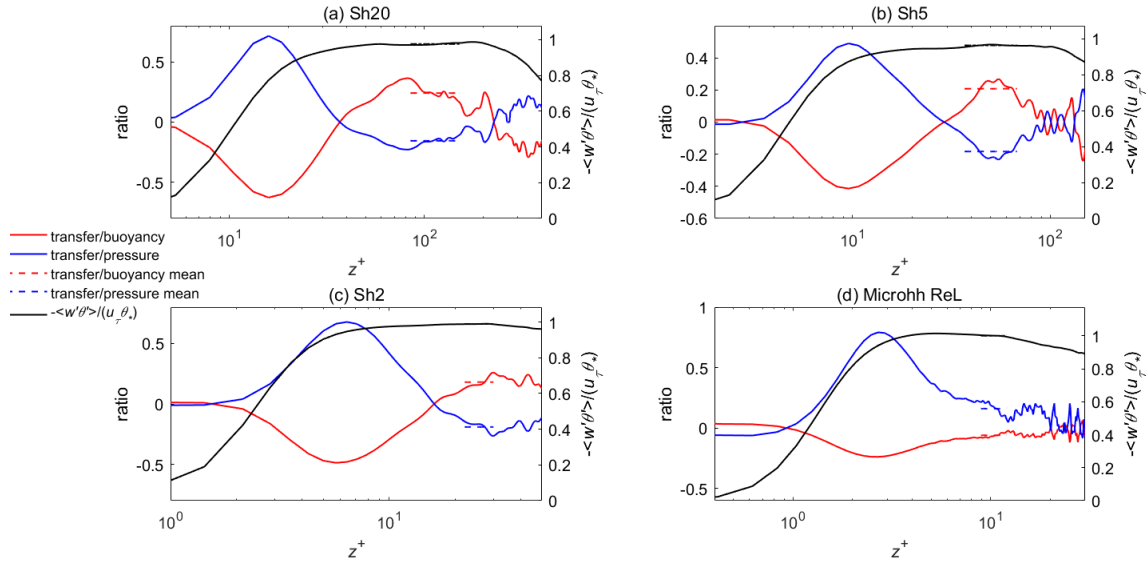

**Fig. S1.** Vertical profiles of the ratio of the transfer term and buoyancy term, the ratio of the transfer and pressure term defined in the transport equation for the heat flux $<w'\theta'>$ (or $\overline{w'\theta'}$), and normalized heat flux in different convective DNS data. The black dashed line denotes the mean $-\frac{<w'\theta'>}{u_\tau \theta_*}$ in the constant heat flux zone. The red dashed line denotes the mean of the ratio of the transfer term and buoyancy term, and the blue dashed line denotes the mean of the ratio of the transfer term and pressure term in the constant heat flux zone.

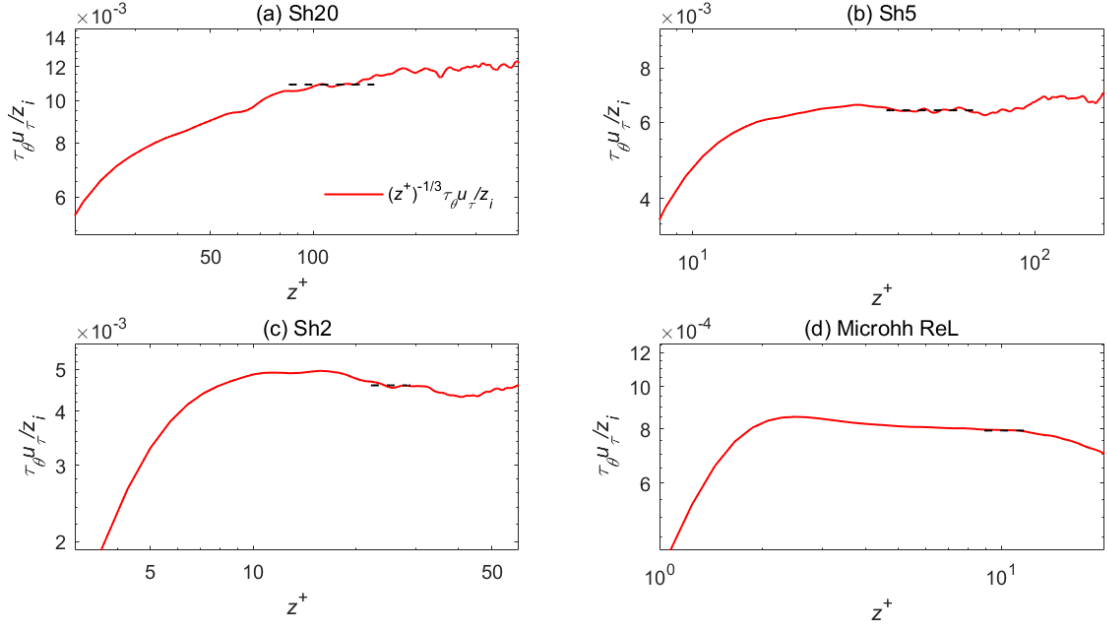

**Fig. S2.** The dimensionless relaxation time $\frac{\tau_\theta u_\tau}{z_i}$ in the $z$ direction in different convective DNS data. $\tau_\theta$ is a return-to-isotropy time scale. The black dashed line indicates $\tau_\theta \propto z^{1/3}$ in the constant heat flux zone.

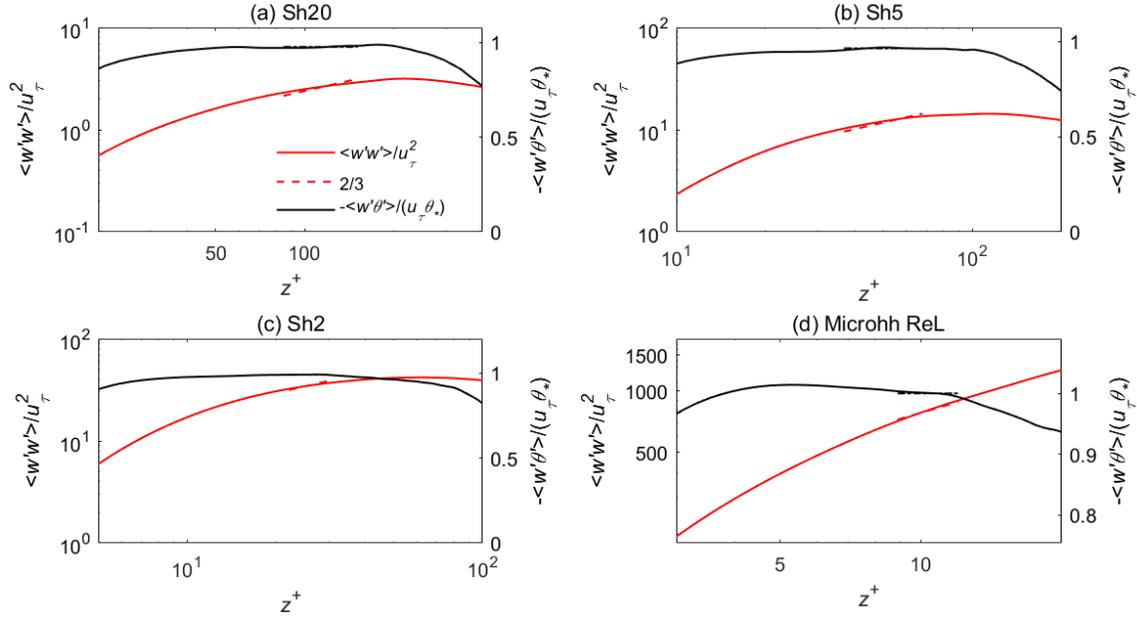

**Fig. S3.** The normalized variance of vertical velocity $<w'w'>$ and heat flux $<w'\theta'>$ in the $z$ direction in different convective DNS data. The black dashed line denotes the constant heat flux. The red dashed line denotes a $2/3$ slope in the constant heat flux zone.

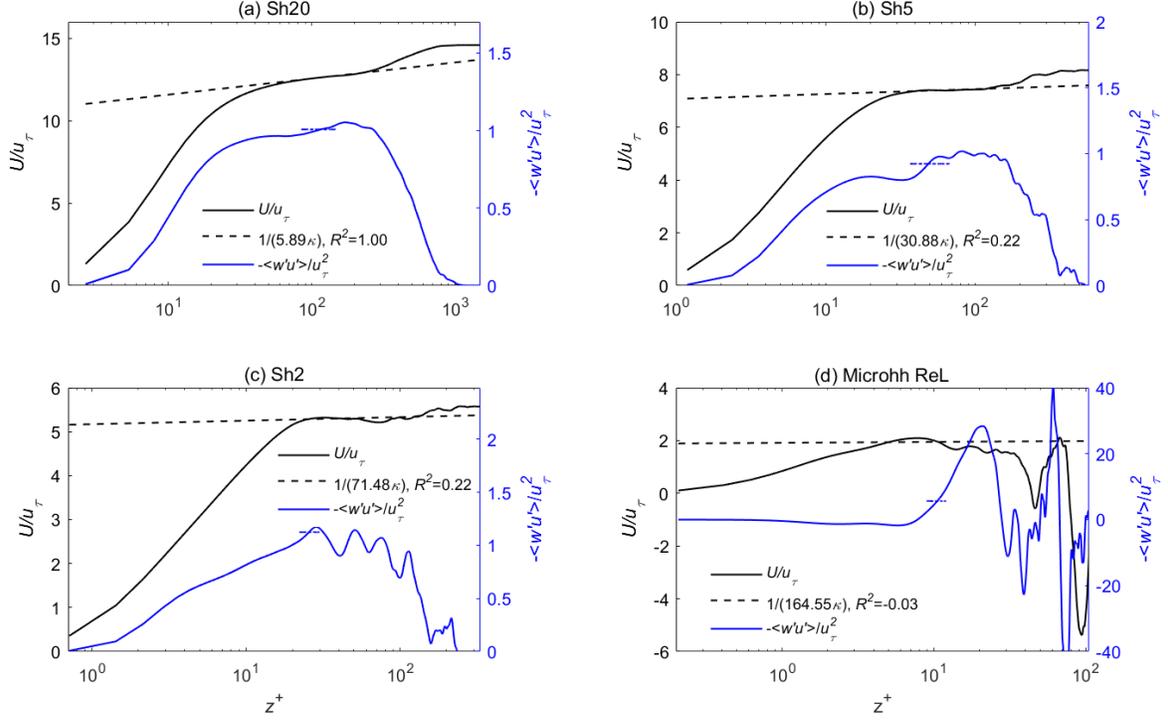

**Fig. S4.** Vertical profiles of normalized streamwise velocity and momentum flux averaged in the $x$-$y$ plane in different convective DNS data. $U$ is mean streamwise velocity in the $x$-$y$ plane, $w'$ is the fluctuation of vertical velocity and $u'$ is the fluctuation of streamwise velocity. The black dashed line denotes the fitted log profile using the slope of the temperature log law. The blue dashed line denotes the mean momentum flux in the constant heat flux zone.

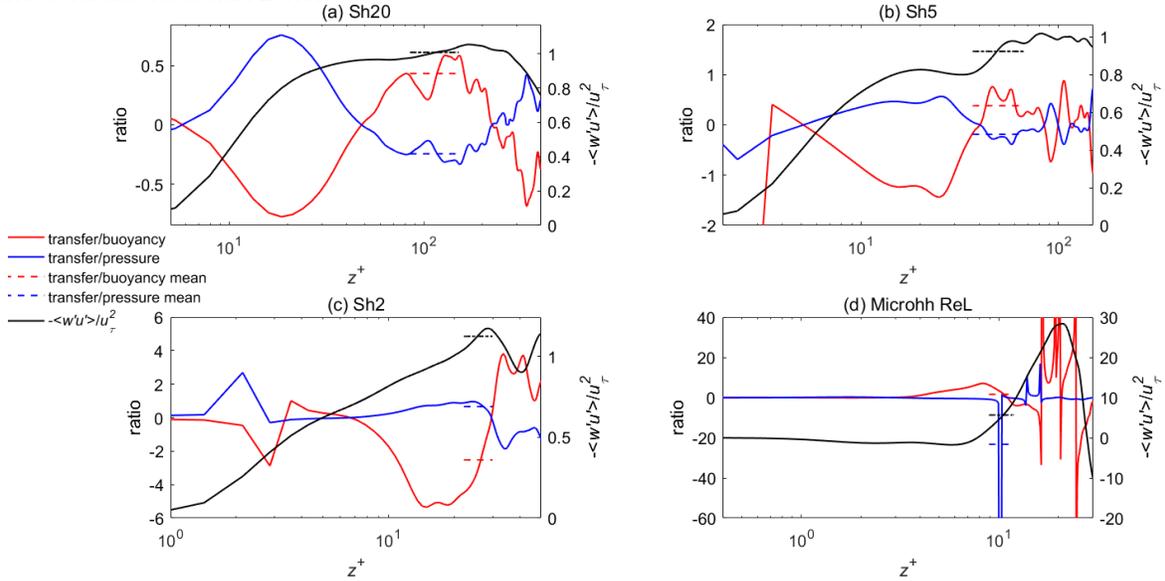

**Fig. S5.** Vertical profiles of the ratio of the transfer and buoyancy term, the ratio of the transfer and buoyancy term defined in the transport equation for the momentum flux $< w'u' >$, and normalized momentum flux $\frac{-<w'u'>}{u_\tau^2}$ in different convective DNS data. The black dashed line denotes mean $-\frac{<w'u'>}{u_\tau^2}$ in the constant heat flux zone. The red dashed line denotes mean of the ratio of the transfer and buoyancy term and the blue dashed line denotes the mean of the ratio of the transfer to pressure term in the constant heat flux zone.

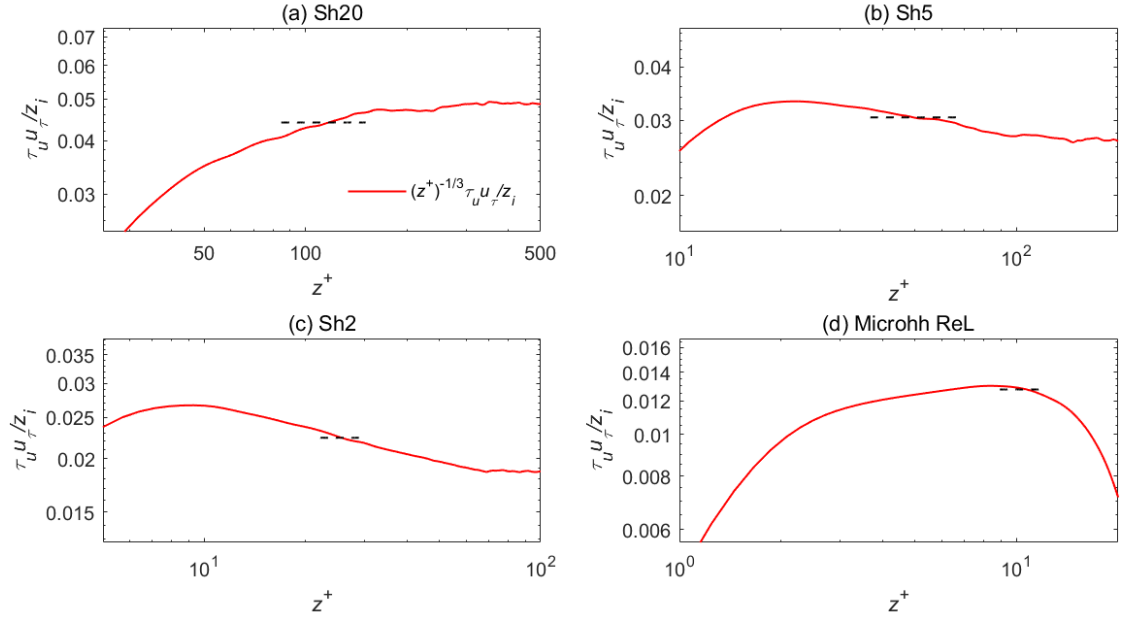

**Fig. S6.** The dimensionless relaxation time $\frac{\tau_u u_\tau}{z_i}$ in the $z$ direction in different convective DNS data. $\tau_u$ is a return-to-isotropy time scale. The black dashed line indicates $\tau_u \propto z^{1/3}$ in the constant heat flux zone.

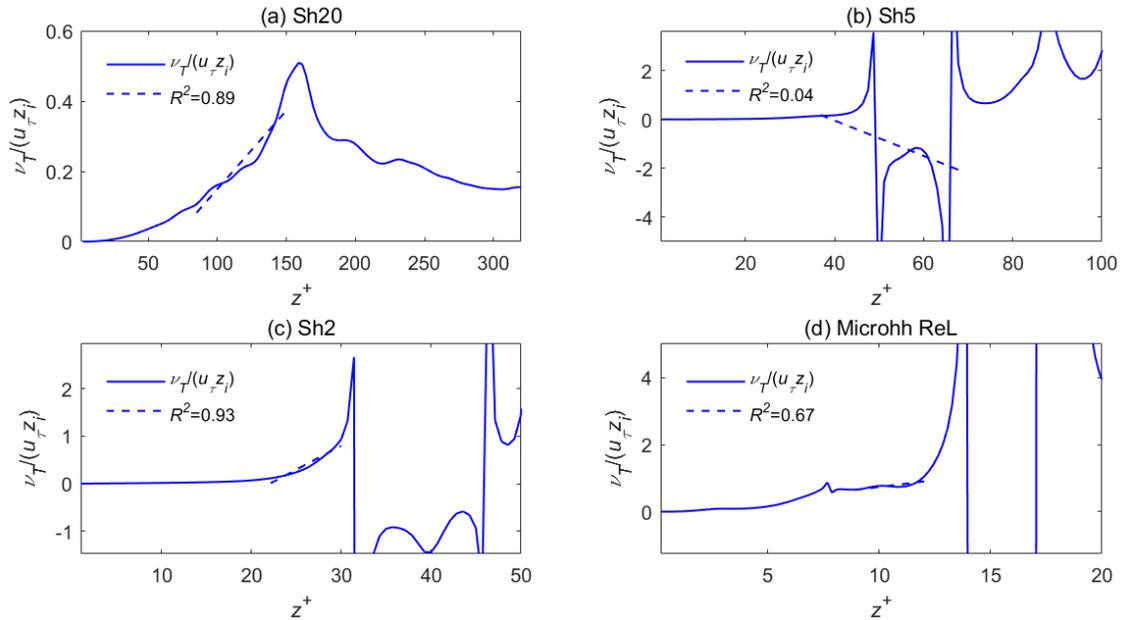

**Fig. S7.** The dimensionless turbulent eddy viscosity $\frac{\nu_T}{u_\tau z_i}$ in the $z$ direction in different convective DNS data. $\nu_T$ is the turbulent diffusivity for velocity. The blue dashed line indicates linear regression in the constant heat flux zone.

**Acknowledgments**
PG would like to acknowledge funding from the National Science Foundation (NSF CAREER, EAR-1552304) and from the Department of Energy (DOE Early Career, DE-SC00142013). The simulations were performed on the computing clusters of the National Center of Atmospheric Research under Project

UCLB0017. We would like to thank Dr. Chiel van Heerwaarden for the help with direct numerical simulation of free convection using MicroHH. We would like to thank Dr. Fred C. Bosveld and Henk Klein Baltink for the help in obtaining the field data at the Cabauw Experimental Site for Atmospheric Research (Cesar) (http://www.cesar-database.nl).

**References**
1. T. von Kármán (1930) Mechanische ahnlichkeit und turbulenz. in *Proceedings of the 3rd International Congress on Applied Mechanics*, pp 85-93.
2. J. L. Lumley, A. M. Yaglom, A century of turbulence. *Flow, Turbulence and Combustion* 66, 241-286 (2001).
3. P. Bradshaw, G. P. Huang, The law of the wall in turbulent flow. *Proceedings of the Royal Society of London. Series A: Mathematical and Physical Sciences* 451, 165-188 (1995).
4. B. Kader, A. Yaglom, Heat and mass transfer laws for fully turbulent wall flows. *International Journal of Heat and Mass Transfer* 15, 2329-2351 (1972).
5. L. D. Landau, E. M. Lifshitz, *Course of Theoretical Physics Vol. 6 Fluid Mechanics* (Pergamon Press, 1959).
6. R. Stull, *An Introduction to Boundary Layer Meteorology* (Kluwer Academic Publishers, Dordrecht, 1988), pp. 666.
7. J.-F. Louis, A parametric model of vertical eddy fluxes in the atmosphere. *Boundary-Layer Meteorology* 17, 187-202 (1979).
8. J. W. Deardorff, Parameterization of the planetary boundary layer for use in general circulation models. *Monthly Weather Review* 100, 93-106 (1972).
9. I. Troen, L. Mahrt, A simple model of the atmospheric boundary layer; sensitivity to surface evaporation. *Boundary-Layer Meteorology* 37, 129-148 (1986).
10. A. Holtslag, B. Boville, Local versus nonlocal boundary-layer diffusion in a global climate model. *Journal of Climate* 6, 1825-1842 (1993).
11. Y. Cheng, M. B. Parlange, W. Brutsaert, Pathology of Monin-Obukhov similarity in the stable boundary layer. *Journal of Geophysical Research: Atmospheres* 110 (2005).
12. A. Monin, A. Obukhov, Basic laws of turbulent mixing in the surface layer of the atmosphere. *Contrib. Geophys. Inst. Acad. Sci. USSR* 151, 163-187 (1954).
13. Q. Li, P. Gentine, J. P. Mellado, K. A. McColl, Implications of nonlocal transport and conditionally averaged statistics on Monin–Obukhov similarity theory and Townsend's attached eddy hypothesis. *Journal of the Atmospheric Sciences* 75, 3403-3431 (2018).
14. C. C. v. Heerwaarden, J. P. Mellado, Growth and decay of a convective boundary layer over a surface with a constant temperature. *Journal of the Atmospheric Sciences* 73, 2165-2177 (2016).
15. A. Obukhov, Turbulence in thermally inhomogeneous atmosphere. *Trudy Inst. Teor. Geofiz. Akad. Nauk SSSR* 1, 95-115 (1946).
16. A. Scagliarini, H. Einarsson, Á. Gylfason, F. Toschi, Law of the wall in an unstably stratified turbulent channel flow. *Journal of Fluid Mechanics* 781, R5 (2015).
17. D. Li, K. Luo, J. Fan, Buoyancy effects in an unstably stratified turbulent boundary layer flow. *Physics of Fluids* 29, 015104 (2017).
18. O. Iida, N. Kasagi, Direct numerical simulation of unstably stratified turbulent channel flow. *Journal of Heat Transfer* 119, 53-61 (1997).
19. S. Sid, Y. Dubief, V. Terrapon (2015) Direct numerical simulation of mixed convection in turbulent channel flow: on the Reynolds number dependency of momentum and heat transfer under unstable stratification. in *Proceedings of the 8th International Conference on Computational Heat and Mass Transfer, ICCHMT 2015*.
20. M. Hölling, H. Herwig, Asymptotic analysis of the near-wall region of turbulent natural convection flows. *Journal of Fluid Mechanics* 541, 383-397 (2005).
21. W. K. George, Is there a universal log law for turbulent wall-bounded flows? *Philosophical Transactions of the Royal Society A: Mathematical, Physical and Engineering Sciences* 365, 789-806 (2007).
22. I. Marusic, J. P. Monty, M. Hultmark, A. J. Smits, On the logarithmic region in wall turbulence. *Journal of Fluid Mechanics* 716, R3 (2013).
23. G. Ahlers *et al.*, Logarithmic temperature profiles in turbulent Rayleigh-Bénard convection. *Physical Review Letters* 109, 114501 (2012).

24. J. W. Deardorff, Preliminary results from numerical integrations of the unstable planetary boundary layer. *Journal of the Atmospheric Sciences* 27, 1209-1211 (1970).
25. S. Pal, M. Haeffelin, E. Batchvarova, Exploring a geophysical process-based attribution technique for the determination of the atmospheric boundary layer depth using aerosol lidar and near-surface meteorological measurements. *Journal of Geophysical Research: Atmospheres* 118, 9277-9295 (2013).
26. P. Seibert *et al.*, Review and intercomparison of operational methods for the determination of the mixing height. *Atmospheric Environment* 34, 1001-1027 (2000).
27. K. McNaughton, R. Clement, J. Moncrieff, Scaling properties of velocity and temperature spectra above the surface friction layer in a convective atmospheric boundary layer. *Nonlinear Processes in Geophysics* 14, 257-271 (2007).
28. J. Laubach, K. G. McNaughton, Scaling properties of temperature spectra and heat-flux cospectra in the surface friction layer beneath an unstable outer layer. *Boundary-Layer Meteorology* 133, 219-252 (2009).
29. H. A. Panofsky, H. Tennekes, D. H. Lenschow, J. Wyngaard, The characteristics of turbulent velocity components in the surface layer under convective conditions. *Boundary-Layer Meteorology* 11, 355-361 (1977).
30. C. Johansson, A.-S. Smedman, U. Högström, J. G. Brasseur, S. Khanna, Critical test of the validity of Monin–Obukhov similarity during convective conditions. *Journal of the Atmospheric Sciences* 58, 1549-1566 (2001).
31. J. W. Deardorff, Numerical investigation of neutral and unstable planetary boundary layers. *Journal of the Atmospheric Sciences* 29, 91-115 (1972).
32. C.-H. Moeng, A large-eddy-simulation model for the study of planetary boundary-layer turbulence. *Journal of the Atmospheric Sciences* 41, 2052-2062 (1984).
33. F. T. Nieuwstadt, P. J. Mason, C.-H. Moeng, U. Schumann, "Large-eddy simulation of the convective boundary layer: A comparison of four computer codes" in Turbulent Shear Flows 8*,* F. Durst *et al.*, Eds. (Springer, Berlin, Heidelberg, 1993), pp. 343-367.
34. J.-P. Mellado, C. Bretherton, B. Stevens, M. Wyant, DNS and LES for simulating stratocumulus: better together. *Journal of Advances in Modeling Earth Systems* 10, 1421-1438 (2018).
35. S. Khanna, J. G. Brasseur, Analysis of Monin–Obukhov similarity from large-eddy simulation. *Journal of Fluid Mechanics* 345, 251-286 (1997).
36. E. Bou-Zeid, C. Meneveau, M. Parlange, A scale-dependent Lagrangian dynamic model for large eddy simulation of complex turbulent flows. *Physics of Fluids* 17, 025105 (2005).
37. H. Schmidt, U. Schumann, Coherent structure of the convective boundary layer derived from large-eddy simulations. *Journal of Fluid Mechanics* 200, 511-562 (1989).
38. S. Pirozzoli, M. Bernardini, R. Verzicco, P. Orlandi, Mixed convection in turbulent channels with unstable stratification. *Journal of Fluid Mechanics* 821, 482-516 (2017).
39. J. P. Mellado, C. C. van Heerwaarden, J. R. Garcia, Near-surface effects of free atmosphere stratification in free convection. *Boundary-Layer Meteorology* 159, 69-95 (2016).
40. A. Haghshenas, J. P. Mellado, Characterization of wind-shear effects on entrainment in a convective boundary layer. *Journal of Fluid Mechanics* 858, 145-183 (2019).
41. K. Fodor, J. P. Mellado, M. Wilczek, On the Role of Large-Scale Updrafts and Downdrafts in Deviations From Monin–Obukhov Similarity Theory in Free Convection. *Boundary-Layer Meteorology* 172, 1-26 (2019).
42. Y. Cheng *et al.*, Deep learning for subgrid-scale turbulence modeling in large-eddy simulations of the atmospheric boundary layer. *arXiv preprint arXiv:1910.12125* (2019).
43. C. C. v. Heerwaarden *et al.*, MicroHH 1.0: a computational fluid dynamics code for direct numerical simulation and large-eddy simulation of atmospheric boundary layer flows. *Geoscientific Model Development* 10, 3145-3165 (2017).
44. A. Apituley *et al.* (2008) Overview of Research and Networking with Ground based Remote Sensing for Atmospheric Profiling at the Cabauw Experimental Site for Atmospheric Research (CESAR)-The Netherlands. in *IGARSS 2008-2008 IEEE International Geoscience and Remote Sensing Symposium* (IEEE), pp III-903-III-906.
45. L. Choma *et al.* (2019) Comparative Analysis of Selected Ceilometers for Practice and Academic Purposes. in *2019 Modern Safety Technologies in Transportation (MOSATT)* (IEEE), pp 35-38.

46. M. d. Bruine, A. Apituley, D. P. Donovan, H. Klein Baltink, M. J. d. Haij, Pathfinder: applying graph theory to consistent tracking of daytime mixed layer height with backscatter lidar. *Atmospheric Measurement Techniques* 10, 1893-1909 (2017).
47. J. Rotta, Statistical theory of nonhomogeneous turbulence. ii. *Z. Physik* 131 (1951).
48. C.-H. Moeng, J. C. Wyngaard, An Analysis of Closures for Pressure-Scalar Covariances in the Convective Boundary Layer. *Journal of the Atmospheric Sciences* 43, 2499-2513 (1986).
49. J. L. Lumley, Computational Modeling of Turbulent Flows. *Advances in Applied Mechanics* 18, 123-176 (1979).
50. J. C. Wyngaard, O. R. Coté, Y. Izumi, Local Free Convection, Similarity, and the Budgets of Shear Stress and Heat Flux. *Journal of the Atmospheric Sciences* 28, 1171-1182 (1971).
51. K. Hanjalić, B. Launder, A Reynolds stress model of turbulence and its application to thin shear flows. *Journal of Fluid Mechanics* 52, 609-638 (1972).
52. G. Willis, J. Deardorff, A laboratory model of the unstable planetary boundary layer. *Journal of the Atmospheric Sciences* 31, 1297-1307 (1974).